\documentclass[preprint, journal]{vgtc}  

\onlineid{1140}

\vgtccategory{Research}

\title{``It's a Good Idea to Put It Into Words'':\\ Writing `Rudders' in the Initial Stages of Visualization Design}

\author{%
  \authororcid{Chase Stokes}{0000-0001-7644-9021}, 
  Clara Hu, and
  Marti A. Hearst
}

\authorfooter{
  \item
  	Chase Stokes and Marti A. Hearst are with University of California, Berkeley.
  	E-mail: cstokes@ischool.berkeley.edu.
}

\abstract{%
  Written language is a useful tool for non-visual creative activities like writing essays and planning searches. This paper investigates the integration of written language into the visualization design process. We create the idea of a `writing rudder,' which acts as a guiding force or strategy for the design. Via an interview study of 24 working visualization designers, we first established that only a minority of participants systematically use writing  to aid in design. A second study with 15 visualization designers examined four different variants of written rudders: asking questions, stating conclusions, composing a narrative, and writing titles. Overall, participants had a positive reaction; designers recognized the benefits of explicitly writing down components of the design and indicated that they would use this approach in future design work.  More specifically, two approaches –- writing questions and writing conclusions/takeaways –- were seen as beneficial across the design process, while writing narratives showed promise mainly for the creation stage. Although concerns around potential bias during data exploration were raised, participants also discussed strategies to mitigate such concerns. This paper contributes to a deeper understanding of the interplay between language and visualization, and proposes a straightforward, lightweight addition to the visualization design process.
}


\teaser{
  \centering
  \includegraphics[width=0.9\linewidth, alt={Alt text.}]{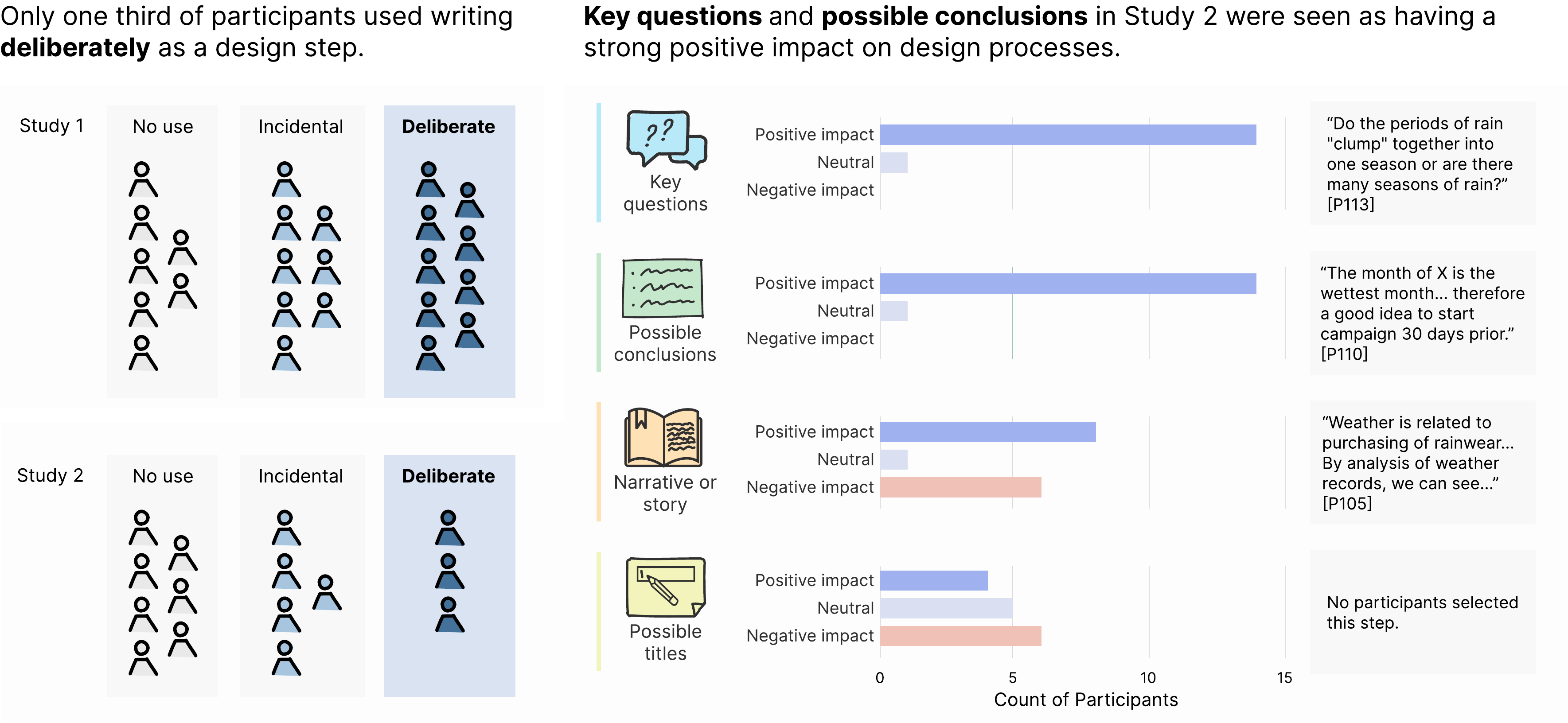}
  \caption{%
  	Main findings from two interview studies. Right: number of participants who currently use writing and with what frequency, in each design step. Both Study 1 and Study 2 found that visualization designers rarely use writing as a concrete design step. Left:  Four types of writing rudders tested in Study 2, participants ratings of each type, and examples of participant-written rudders. 
   Of the variants proposed in Study 2, key questions and possible conclusions were seen as the most beneficial. 
  }
 
  \label{fig:teaser}
}

\graphicspath{{figs/}{figures/}{pictures/}{images/}{./}} 

\usepackage{mathptmx} 
\usepackage{enumitem}
\usepackage{fontawesome}
\usepackage{colortbl}
\usepackage{url}
\usepackage{float}
\usepackage{pifont}
\usepackage{graphicx}

\newcommand{\pheading}[1]{\vspace{2px}\noindent\textbf{#1}}
\newcolumntype{R}[1]{>{\raggedright\arraybackslash}p{#1}}
\newcommand{\centeredmulticolumn}[3]{\multicolumn{#1}{c}{\raisebox{-2pt}{\includegraphics[height=9pt]{#2}} #3}}

\arrayrulecolor{gray} 

\newenvironment{tightItemize}{\begin{itemize} \itemsep
-2.5pt}{\end{itemize}}

\usepackage{tabularx}
\usepackage{tabu}                      
\newcolumntype{P}[1]{>{\arraybackslash}p{#1}}

\usepackage{xurl}

\usepackage{booktabs}

\begin{document}



\maketitle

\section{Introduction}
\label{sec:intro}

The process of visualization design puts great attention on the creation of visual elements. Activities that produce visual artifacts, such as sketching and wireframing, are common steps in visualization design \cite{parsons2021understanding}.
By contrast, taxonomies and studies of design practice have not focused on the use of \textit{writing} in the design of visualizations. 
For creative tasks outside of data visualization, taking time to write down a plan or outcome is a standard practice. For example, library science has long advised writing down the information need as a prelude to effective search \cite{wilson2004talking,ross2019conducting}. The introduction to Russell's \textit{Joy of Search} \cite{russell2019joy} advocates this approach for web search:  
\begin{quote}
 ``I know, it sounds too simple to actually work.  But if you take ten seconds to write down your question BEFORE you start ... you'll find your research process will be much, much more effective.'' (p. 7)
\end{quote}
\noindent
Research on essay writing instruction finds benefits in  pre-writing steps, including writing outlines, lists, notes, or concept webs \cite{graham2007meta}.
In visual endeavors such as animation and film, scripts and screenplays provide a narrative foundation for eventual visual output. This paper explores this overlooked aspect of design for the creation of visualization.

The origin of this paper stems from the direct experience of the authors in creating visualizations to serve as stimuli for a research study. At first, we struggled to determine what the narrative of the visualization would be and how we should write the accompanying text. We took a step back from the visualization itself to write a short paragraph describing the visualization and the story of the data. After writing this narrative, the design of the visualization was more successful. 

In this paper, we focus on the use of language to support the framing or the narrative of the visualization. We use the term \textbf{rudder}, which refers  to a mechanism to steer a boat, as well as, more metaphorically, ``a guiding force or strategy'' \cite{mw:rudder}.
A written rudder provides direction in the design of the visualization and helps to maintain focus on the message and goals of the project. 
Written rudders are also flexible, just as a boat's rudder may be pivoted to move the boat in a new direction. 
The purpose of a rudder is to \textit{guide} the design, similar to how a sketch acts as a starting point for determining visual representations. 

Design practice currently uses writing in  ways that differ from the rudders notion.
Designers may use writing for user research, such as reviewing or summarizing insights from user interviews. 
It is also a common practice to document specific design decisions or write down design requirements \cite{munzner2009nested, walny2019data, mckenna2014design}. 
User interviews provide data on the insights to display but this data does not itself describe the design's message or goals. 
Design documentation is about tracking and justifying choices made throughout the design development, while the writing rudder serves as a guide before or during the design process.

Design requirements are the most similar to a writing rudder, but there are a few areas where the two diverge. Design requirements typically involve technical specifications as well as possible user needs or objectives. They may also contain limitations for the project, and the specific data that will be used. Writing rudders, on the other hand, focus primarily on guiding (rather than requiring or instructing) the design process and focus on the message or story of the design.

The idea behind the writing rudder can also be situated in the context of design philosophy. As an element of the design process, a rudder can impose a ``discipline,'' acting as a useful constraint to focus and guide the iteration of ideas and design \cite{schon1987educating}. Writing rudders also support the iterative nature of design, since they are lightweight and so can be recreated or adjusted as the design evolves or parameters change. 

We investigate two main research questions. First, \textbf{how do designers currently use writing during the design process?} 
Second, \textbf{what is the perceived impact of writing rudders on the design process?}
In this paper, we use the following terms and definitions: 

\begin{tightItemize}

\item \textbf{Visualization designer}: a professional who creates visual representations of data as part of a paid role, typically for a specific ask or objective. Also referred to as  \textbf{designer} or \textbf{practitioner}. 
\item \textbf{Design process}: the dynamic, iterative set of activities undertaken while creating visual representations from raw data. 
\item \textbf{Writing rudder}: 
hand-written or typed language created and/or used during the design process, describing the message, story, or key goals of the design itself. Also referred to as a \textbf{written rudder} or \textbf{rudder}. A \textbf{rudder variant} or \textbf{variant} is a specific form of this language (e.g., questions the design may address).
\item \textbf{Text elements}: typeface features on visual representations of data (e.g., captions, annotations, etc.).

\end{tightItemize}

\smallskip

\pheading{Contributions:}
We contribute two interview studies with visualization designers which illustrate the potential impact of writing rudders.  

Study 1 finds that the deliberate use of writing as a design step among visualization designers is uncommon. When writing is used, it primarily informs the designer's understanding of the visualization's goals and the generation of creative ideas. We conclude that writing is an underutilized step in visualization design. 

Study 2 builds upon these findings by exploring the tangible benefits and drawbacks of incorporating specific written rudder variants, shown in \cref{fig:teaser}, into the visualization design process. Four variants were proposed, two of which were found to be especially beneficial to the beginning stages of the design process. The written artifacts created from these steps may also have important evaluative uses. 

This research supports the utility of writing down guiding text as part of the design process and suggests additional areas for future work and evaluation. 

\section{Related Work}
\label{sec:rw}

\subsection{Practical Insights from Visualization Design}

Academic research often lacks actionable insights for practitioners, indicating a need to bridge the gap between research questions and design practices \cite{smits2023towards, parsons2020methods}. In the broader field of HCI, there has been a shift towards practice-oriented research programs to enhance practical relevance of frameworks and theories \cite{kuutti2014turn, goodman2011understanding}. 
In visualization design studies, researchers engage with design practices in HCI, collaborating with domain experts to address problems through visualization systems \cite{sedlmair2012design, meyer2019criteria}.
These systems can contribute both practical solutions and theoretical insights. 
There is ongoing discussion about the dissemination and relevance of design research findings \cite{gray2014bubble, parsons2021understanding}.

The design process is nonlinear and iterative \cite{parsons2021understanding, parsons2020methods, parsons2021fixation, alspaugh2018futzing}. Design involves insights from science and research, such as visual encodings and marks/channels, as well as artistic sensibility;  visualization designers often incorporate elements of creativity into their designs. Designers often use guidelines, heuristics, or examples of past designs to assist in ideation and to avoid fixating on a particular design or approach \cite{parsons2020methods, parsons2021fixation, bako2022understanding}. Designers require flexibility in their approaches depending on the context, audience, and particulars of the data being displayed \cite{bigelow2014reflections, walny2019data}. This need is even more salient in environments with sensitive and/or important data, such as the COVID-19 pandemic \cite{zhang2022visualization}.

There have been a few studies examining the use of written guides in visualization design. Lee-Robbins et al. \cite{lee2021learning} found that participants selected more effective visualizations when provided clear learning objectives (e.g., ``Spot outliers within the data.''). Learning objectives are similar to the concept of written rudders that we explore in this paper. However, learning objectives take specific forms and contain certain information, while rudders are more flexible in form and content. 

Some researcher-designed tools consider the use of natural language in the visualization design process. For example, InkSight \cite{lin2023inksight} allows visualization creators to augment their iterative sketching practices with generated data facts within a computational notebook environment. This method of exploratory data analysis supports concrete insights through natural language as well as freeform investigations through sketching.
Storyboarding frameworks have also proved useful for the design of interactive systems, including visual analytics \cite{walker2015storyboarding, truong2006storyboarding}. Text accompanying the storyboard was particularly useful for understanding the overall narrative \cite{truong2006storyboarding}.

\subsection{Frameworks in Visualization Design}

Researchers have created a variety of frameworks to capture key steps in the design process \cite{sedlmair2012design, munzner2009nested, mckenna2014design, wodehouse2010integration, mccurdy2016action}. These frameworks tend to incorporate similar stages of the design process: a stage for understanding the data and the overall context, a stage for generating ideas to show the data, and a stage for creating the design.  
While design frameworks are helpful for considering the different steps and actions that make up visualization design, none explicitly incorporate writing in any stage.

In this paper, we use the Design Activity Framework (DAF) as our main framework \cite{mckenna2014design}. In comparison to other models and frameworks, the DAF fits most clearly with the nonlinearity of design practices and is the most generalizable across design contexts. The DAF provides four overlapping activities in the design process: understand, ideate, make, and deploy. Similar to the nested model \cite{munzner2009nested}, each stage typically provides an output for development in the following stage. However, there is more overlap and less linearity in the DAF process. Designers typically first \textit{understand} the users, their data, and the project context. They then \textit{ideate} different ways of communicating key information (e.g., different visual encodings) using sketches or low-fidelity prototypes \cite{bigelow2014reflections, parsons2021fixation, fallman2003design}. These lo-fi ideas are then \textit{made} into tangible prototypes and are \textit{deployed} when the final design is decided and created in full. These areas are intricately linked but also separable for individual study.

\subsection{Teaching Visualization Design}

Teaching the design of visualizations involves not only imparting knowledge of tools and techniques but also fostering an understanding of the conceptual underpinnings that guide the effective communication of data. The teaching of visualization principles typically involves an understanding of visual encodings \cite{cleveland1984graphical},  communicating data clearly through data storytelling \cite{knaflic2015storytelling, tufte1985visual, munzner2014visualization}, and instruction in tools for visualization design, such as Tableau or d3 \cite{beyer2016teaching}. This often takes the form of hands-on projects and exercises to encourage the application of these teachings in real-world scenarios \cite{kammer2021experience, bach2023challenges}.
Also recommended are design and redesign activities \cite{Viégas_Wattenberg_2015}, peer critique \cite{beasley2020leveraging}, and design thinking \cite{mckenna2014design, wassink2009applying}.
Worksheets and other writing activities can formalize some of these steps, requiring the student to think through specific steps in creating and conceptualizing a visualization \cite{byrd2021activity}.

\subsection{Narrative Visualizations}

The impact of language in visualization has recently become a growing area of interest in visualization research  (e.g,. \cite{stokes2021give, ottley2019curious,borkin2015beyond,kong2018frames, wanzer2021role,ajani2021declutter}). Data stories and narrative visualizations often incorporate language elements alongside data, animation, and other visual elements \cite{segel2010narrative, hullman2011visualization, kosara2013storytelling, lee2015more, hulllman2013deeper, figueiras2014narrative}. While these elements do not always result in a high degree of engagement, the use of data storytelling methods can increase overall comprehension for certain tasks \cite{boy2015storytelling, shao2024data}.

In addition to being important for data communication overall, data storytelling and narrative infographics are increasingly recognized as crucial components of visualization design education \cite{hearst2017teaching}. The role of narratives or stories in data visualization allows for the development of broader skills alongside visualization techniques \cite{bach2023challenges, parsons2023preparing}. 

There are systems designed to support storytelling with data \cite{satyanarayan2014authoring}, including those that combine text and visual components of the design \cite{srinivasan2018augmenting, latif2021kori, chen2020supporting}. One recent contribution to this space, Epigraphics, explores a similar idea to writing rudders \cite{zhou2024epigraphics}. Users of Epigraphics begin with text as a first-class object, writing the key message they plan to convey. This is an example of a written rudder embedded within a tool for designing infographics. In this study, we focus on the use of rudders for visualization design broadly and explore additional variants. 
Outside of this example, the creation of a narrative visualization centers more on how text elements are used in the resulting visualization, rather than the use of language \textit{during} the design process.

Writing rudders also share similarities with the concepts of `model checking' and graphical belief elicitation \cite{kale2024evm, mahajan2022vibe, kim2017explaining}. By generating pseudo-hypotheses about the trends or features in the data, analysts can more effectively evaluate the accuracy of these hypotheses. 
Similarly, writing rudders prompt designers to consider the audience’s perspective and anticipate what conclusions might be drawn from the visualization depending on the actual appearance of the data. 
This use of rudders encourages designers to externalize their assumptions and consider the audience’s perspective, enhancing the overall design process in a way similar to how belief elicitation can enhance analysis and interpretation.

We mention narrative visualizations and text elements to distinguish them from the focus of this paper, in which we study the use of written language in the design process.
There are instances where the concepts can be connected. For example, a designer working on a narrative visualization may use a written rudder to illustrate the kind of conclusions readers might come to after viewing the display. Additionally, a designer could use a written rudder to influence the title or caption that ultimately appears in their final design. Despite the ties between the ideas, written rudders are a distinct contribution which focuses on the use of writing in the visualization design process, bringing in the affordances of language to guide visual design.

\section{Study 1}
\label{sec:study1}

In the first study, we conducted semi-structured interviews with visualization designers to shed light on the following research question:

\pheading{RQ1}: How do designers use writing during the design process?
 
\begin{table}[h!]
\centering
\def\arraystretch{1.1}
\begin{tabular}{l|c|c}
\hline
\addlinespace[2pt]
 \textbf{Demographic} & \textbf{Study 1} & \textbf{Study 2} \\
  & n = 24 & n = 15 \\
  \addlinespace[2pt]\hline\addlinespace[2pt]
   \multicolumn{3}{l} {\textit{Years of Experience}}  \\
 \quad 1-3 years & 3 & 4 \\
 \quad 4-6 years & 9 & 4 \\
 \quad 7-9 years & 7 & 2 \\
 \quad 10+ years & 5  & 5 \\\addlinespace[2pt]\hline\addlinespace[2pt]
 \multicolumn{3}{l}{\textit{Time Spent Designing (per week)}}  \\
 \quad Less than 5 hours & 2 & 2 \\
 \quad 5-10 hours & 5 & 5 \\
 \quad 11-20 hours & 7 & 2 \\
 \quad 21-30 hours & 5 & 2 \\
 \quad 30+ hours & 5 &  4 \\
\addlinespace[2pt]\hline\addlinespace[2pt]
 \multicolumn{3}{l}{\textit{Industry Sector}}  \\
 \quad Broadcasting/Journalism & 5 & 3 \\
 \quad Manufacturing & 3 & 0 \\
 \quad Medical/healthcare & 4 & 1 \\
 \quad Non-profit/Government & 3 &  4 \\
 \quad Research & 2 & 5 \\
 \quad Scientific or Technical Services & 4 & 1 \\
 \quad Software & 3 & 1 \\
\addlinespace[2pt]\hline
\end{tabular}
 \caption{Information about participants' experience and work context. 
 }
 \label{tab:study_participants}
\end{table}

\subsection{Methodology}
\label{sec:study1_method}

Participants completed a 5-minute pre-interview survey about their work with data visualizations, followed by a 60 minute semi-structured interview on Zoom. The interview was recorded and automatically transcribed by the Zoom Cloud Service. They were compensated with a \$30 Amazon gift card. 

The pre-interview survey contained questions regarding features of the participants' work environments, such as the time per week they spend creating or working with data visualizations, the size of their company, and the number of people they frequently collaborate with. It also collected details about their work with visualizations, including the types of visualizations they create and the tools they frequently use, as well as information, such as the participants' level of experience, their gender, and their employment status. 

Interviews began with an overall discussion of the participant's role and responsibilities in data visualization design, continuing on to discuss different aspects of their design process, such as tools used, collaboration practices, and the use of writing in the design process. When possible, the participant and interviewer would discuss an example design to ground this discussion, selected by the participant prior to the interview. In some cases, this was not permitted by the participant's company. This portion of the interview took approximately 25 minutes.

The rest of the interview examined implementation of text elements (e.g., annotations) in visualizations, which is a separate topic intended for a different set of research questions. This interview protocol allowed us to optimize participant time, but the primary focus of this paper is on the design process of visualizations, rather than the use of text elements. For this analysis, we only focus on content from the interviews that pertained to the design process and the use of writing.
The interview protocol can be found in supplementary materials. Participants are referred to with their assigned ID number [P\#].

\subsubsection{Participants}
\label{sec:study1_participants}

24 visualization designers participated in this study. Participants were recruited from the \href{https://www.datavisualizationsociety.org/}{Data Visualization Society} (DVS), \href{https://newsnerdery.org/}{News Nerdery}, and public posts to X (Twitter) and LinkedIn. Recruitment materials had an emphasis on recruiting data journalists in addition to more traditional visualization designers to account for the possibility of their unique perspectives on the integration of text and visualization. All participants, regardless of their primary role, were involved in visualization design work.
To be eligible for this study, participants were required to be based in the United States, be fluent in English, and spend a portion of their time at work designing or creating visualizations. These eligibility requirements were aimed to provide some consistency across participants and ensure study-relevant experience.

22 of the participants were employed full time, with one participant employed part time and another unemployed, looking for work. A majority of participants were women (14), with eight men, one non-binary person, and one trans man. 
Most participants worked in medium-sized (7) or enterprise-sized companies (11).
Further information on participants can be found in \cref{tab:study_participants}, and supplementary materials.

\subsubsection{Coding}
\label{sec:study1_coding}

Interviews were coded along three dimensions: the design outcome, writing in their design process, and the specific phase of the design process during which writing was used, when relevant. 
Descriptions of each code can be found in the codebook in supplementary materials.

The codes for the \textbf{outcome} dimension were developed post-interview, derived from the common outcomes mentioned by participants in the interviews via grounded coding. While these codes could be further refined during the coding process, no new codes were added. This code served to provide context for participant practices. 

We also developed a separate dimension to address our main research question. While we knew we would code a dimension for \textbf{writing} prior to conducting the interviews, the precise codes (No use/none, Incidental, and Deliberate) were developed through the coding process. 
Distinguishing between incidental and deliberate use of writing required consideration of participants' practices with written elements. Specifically, we focused on the regularity and impact of the written elements on the design process. 
For example, deliberate use was identified when participants systematically created and referred to written notes as a main step in their design process. Incidental use, on the other hand, was identified when such notes were created sporadically and had minimal impact on the design's development. 
It was challenging to make this subjective judgement from interview transcripts; we aimed to mitigate this by using detailed coding guidelines and seeking consensus among coders when assigning these labels.

Additionally, we accounted for the \textbf{stage} of the design process that the rudders impacted. Stage codes (Understand, Ideate, Make, Deploy) were drawn from the Design Activity Framework (DAF) \cite{mckenna2014design}. Prior to conducting the interviews, we considered different design frameworks. The DAF was chosen for its compatibility with other frameworks \cite{munzner2009nested, mccurdy2016action} and its separate but intersecting stages of design. 

The first and second authors engaged in detailed discussions about the dimensions and codes, applying them to two transcripts. The codebook was then refined for improved clarity. Following this, the two coders independently coded the entire set of 24 transcripts. They met to discuss discrepancies and reach consensus on disagreeing codes. 
For the codes where consensus was not found, the third author was brought in as a tiebreaker to review the relevant transcripts, without prior knowledge of the codes applied by the first two coders. After considering responses from all three coders, consensus was reached on all coding categories.
Interrater reliability (Cohen's kappa) for the two original coders was calculated for all codes and can be found in supplemental materials. There was moderate to substantial agreement between coders, with kappa values ranging from 0.66 to 0.75.

\subsection{Results}
\label{sec:study1_results}

We distilled two themes regarding the use of writing in participants' design processes. Overall, writing was not frequently used as a deliberate and intentional part of the design process, as shown in \cref{fig:teaser}. 
When participants did use writing (deliberately \textit{or} incidentally), this tended to occur in the early stages of designing. 

\subsubsection{Theme 1: Use of Writing is Relatively Uncommon}

Participants tended to use visual methods in their design processes. Some participants (8/24) began their visualization process with sketching, and many participants (17/24) used sketching as at least part of their design process.
For example, P19 said that while on an initial call with a client, they would ``just sketch out the chart... just quickly take my pen and sketch out. Other times, if it's more complex, I'll draw a more complex chart in my notes.'' Across participants, this sometimes took the form of paper and pencil but could also occur using digital tools like Figma. P11 commented on the fast-moving landscape of these tools, ``So we started out with Axure and Illustrator... now we are on Figma. God knows what we will be on next month.''

Other participants (8/24) began their design process by putting the data directly into a tool to explore and visualize. 
The initial use of a visualization tool versus sketching could depend on the complexity of the data, as stated by P18, ``I'll try to sketch it out or just mock something up... to try to get an idea of what something is gonna look like. Or I'll take Tableau, Power BI and just throw the data in there, see what happens. And then start refining if it's less complicated.''
Putting the data directly into the tool facilitated speed and ease in the design process: ``it's just easier to test different chart types that we're looking at'' [P14]. The speed of changing chart types and variables in a tool often meant that, at times, it would take longer for the participant to draw the chart than it would to make the same chart in Tableau.
The use of these tools also supported exploratory data analyses (EDA).

For most participants (15/24), writing was either not used at all (7/24) or not used as a distinct part of the design process (8/24). In the latter case, participants would mention taking notes or having written documents, but these were not integral to their design process. Around a third of participants (9/24) used writing in a deliberate way
(more detail in \cref{fig:teaser}).
Only two participants \textit{started} their design with some form of writing similar to a written rudder. For both designers, this took the form of a headline or written report from another collaborator. 

Participants (7/24) who did not incorporate writing \textit{at all} during their design process tended to think of the process as more internal. P1 stated, ``I think it's happening internally. I don't list out key takeaways ... I guess it comes up in the process.'' In some cases, despite explicit probing from the interviewer, written rudders were never mentioned.

Participants (8/24) who \textit{incidentally} incorporated writing did not consider it an important or consistent part of their design process. For example, P18 mentioned that they ``try to keep notes as I'm going, cause as I'm coming up with a design, or really, as I'm working on it, I'll just have random stuff pop on my head.'' In this case, the use of language was sporadic and informal, serving primarily as a tool for capturing fleeting ideas rather than a formal element of designing.

Participants (9/24) who \textit{deliberately} incorporated writing in their design process tended to think of them as intentional steps, often mentioning without probing from the interviewer.  This could take the form of a formal document, such as a ``written and approved strategic plan'' [P3] or ``paragraphs that are really data heavy'' [P13]. 
In exploratory reviews of the codes, journalists (5/5) and participants who created text and visual reports (7/10) used writing in a deliberate way. However, all journalists created text and visual reports, thus conflating the values. 

Overall, the use of writing in the design process
was relatively uncommon in our sample, with only about a third or participants considering it a pivotal or concrete step in their design process. 

\subsubsection{Theme 2: Language Used in Early Design Stages}

For the participants who used writing in any minor capacity (17/24), this step most frequently played a role in the \textit{understand} stage of the design. 
In general, these writings set the scene for what the visualization addressed and the specific needs served.

Participants (11/15) relied on this step as a way to better understand the ideas behind the design and key questions the design would be used to answer. P20 described their use of written language as, ``After I have that initial conversation, I like write it all up. This is the question. This is the context. This is the data we're going to use. This is how we think we're going to communicate it.'' In other words, the preparation for the design is written out, with key questions and data attributes captured in concrete language prior to beginning the design. 
Another participant created a short description of `` what the graphic is supposed to show, which is usually two sentences'' [P17]. 

In data journalism contexts, the actual written artifact may come from sources other than the participants themselves: ``I request [the story draft] because it's helpful for me... I'm less likely to make a mistake if I see the whole story, even if it's just reporter notes'' [P6]. The language  provides information on the problem domain, but it comes from a source other than the designer. However, in cases where journalists were writing their own reports, a similar process took place where the draft was written first, and data-heavy paragraphs were replace with ``preliminary charts'' for review [P13].

Writing was  also used during the \textit{ideate} stage of design (7/15), where participants were brainstorming different ways to address the needs of the design. For example, after finalizing the goals and intents of the design, P23 has ``a whole notes document going of things that just occur to me.''

\section{Study 2}
\label{sec:study2}

\begin{figure}
 \centering
 \includegraphics[width=0.88\linewidth]{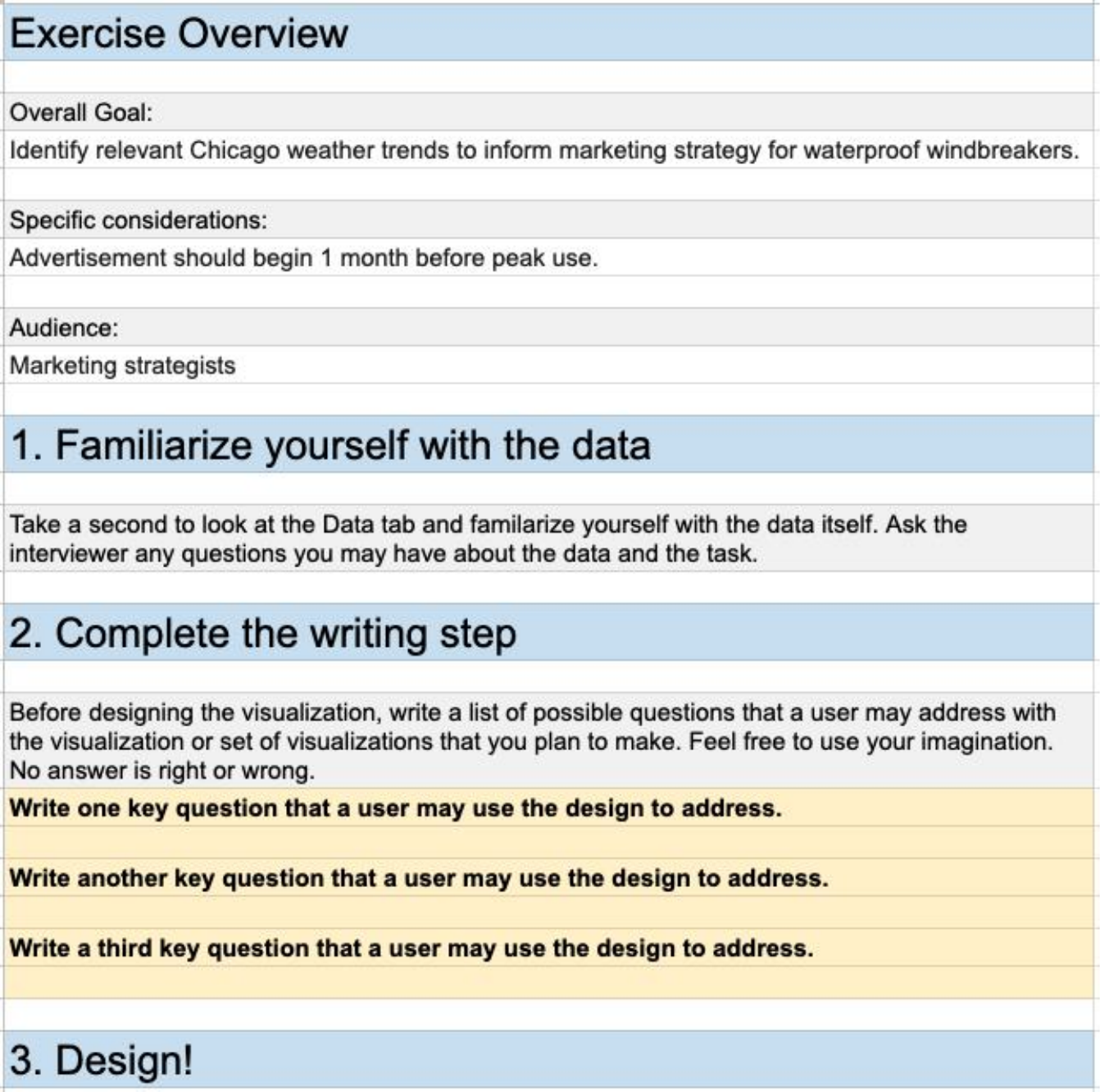}
 \caption{Template for the design exercise in Study 2. The ``writing step'' was filled in with the rudder variant selected by the participant. An example is shown here for ``User's key questions.''}
 \label{fig:exercise_template}
 \end{figure}

Following the findings from Study 1, we faced several open questions about the impact of writing  on designer workflows. Is writing beneficial for designers? What form should these steps take? Which stages of the design process do they impact?

We extracted four different ways that people might use written language from the interviews in Study 1 in order to evaluate them with designers.  
We conducted a second set of semi-structured interviews to collect feedback on writing rudders with a new set of visualization designers. This included a brief design exercise and a post-interview survey to  address the second research question:

\smallskip
\pheading{RQ2}: What is the perceived impact of writing rudder variants on the design process?

We focused primarily on the design \textit{process}, rather than the design \textit{outcome}.
As a preliminary study, the primary aim was to evaluate the feasibility and interest in interventions like written rudders. During pilot interviews for this study, designers expressed concern that observing their design process or judging their outcomes could introduce pressure to the conversation with the interviewer, similar to a job interview. In an effort to avoid this connotation and to emphasize the evaluation of the rudder, rather than the design abilities of the participants, the study was scoped to focus on the process, encouraging the participants to share their true impressions of written rudders. 

Based on insights from Study 1, we focused this study specifically on the beginning stages of the design process. The initial design stages, particularly the \textit{understand} and \textit{ideate} phases, were where participants most frequently utilized writing to shape their designs. By concentrating on these stages, we aimed to capture the most impactful and relevant use of writing in the design workflow, ensuring that our findings are both practical and applicable to real-world scenarios.

\subsection{Methodology}
\label{sec:study2_method}

\begin{table*}[ht]
\centering
\def\arraystretch{1.4}
\begin{tabular}{llllll}
\hline
\textbf{Rudder Variant} & \textbf{Getting Started} & \textbf{Understand} & \textbf{Ideate} & \textbf{Make} & \textbf{Overall} \\
\hline
\raisebox{-2pt}{\includegraphics[height=9pt]{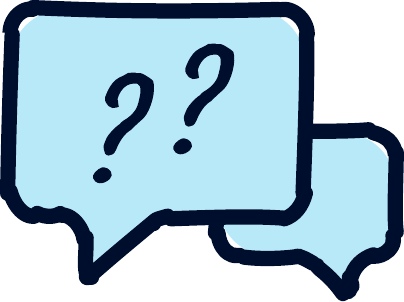}} Questions   & 4.13 (0.74) & 4.53 (0.64) & 4.27 (0.46) & 4.33 (0.82) & 4.30 (0.68) \\

\raisebox{-2pt}{\includegraphics[height=9pt]{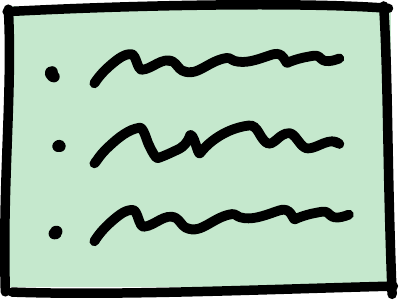}} Conclusions & 4.20 (0.86) & 4.53 (0.64) & 3.93 (0.80) & 4.13 (0.83) & 4.20 (0.80) \\

\raisebox{-2pt}{\includegraphics[height=9pt]{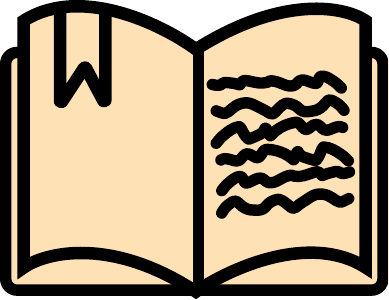}} Narratives  & 3.33 (1.29) & 3.53 (1.06) & 3.40 (1.18) & 3.47 (1.06) & 3.43 (1.13) \\

\raisebox{-2pt}{\includegraphics[height=9pt]{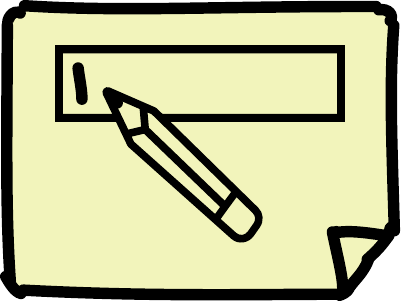}} Titles & 2.73 (1.33)  & 3.00 (1.20) & 2.93 (1.28) & 2.87 (1.13) & 2.88 (1.21) \\
\hline
\end{tabular}
\caption{Average ratings for each rudder variant for the different stages of visualization design, as well as the average rating overall. Standard deviations are included in parentheses. 
Ratings were made on a scale from 1 (negatively impact) to 5 (positively impact).  
}
\label{tab:avg_ratings}
\end{table*}

Participants completed the same pre-interview survey as described in  Study 1. This was followed by a 60 minute semi-structured interview on Zoom. The interview was recorded and transcribed by the Zoom Cloud Service. After the interview, participants completed a post-interview survey and were compensated with a \$30 Amazon gift card. 

There were four distinct phases in the interview: designer role, introduction of variants, design exercise, and reflection. 
Interview materials can be found in supplementary materials.
As in Study 1, interviews began with a broad discussion of the participant's role in designing visualizations. The interview focused the conversation specifically on the beginning of the design process, attempting to capture the initial steps a designer typically takes.
This design exercise was piloted by three graduate students, five participants from Study 1, and six additional participants recruited for Study 2. The pilot studies helped to shape the scope of the data, the time frame, and the description of the design task. 

After asking about the participant's design process and context, the interview moved to the next phase, where the interviewer introduced four possible written rudders that could be incorporated into the early stages of design, prior to bringing the data into a visualization tool. Specifically, these were: 

\begin{tightItemize}
  \item \raisebox{-2pt}{\includegraphics[height=9pt]{figs/questions_icon.pdf}} \textbf{Key questions}: Write down the key questions that a user/reader may use the visualization(s) to address. 
  \item \raisebox{-2pt}{\includegraphics[height=9pt]{figs/conclusion_icon.pdf}} \textbf{Possible reader conclusions/takeaways}: Write down ideas for possible conclusions readers might make when viewing the visualization(s).
  \item \raisebox{-2pt}{\includegraphics[height=9pt]{figs/story_icon.pdf}} \textbf{Narrative/story}: Write down a brief story that conveys the main points the visualization(s) might express.
  \item \raisebox{-2pt}{\includegraphics[height=9pt]{figs/title_icon.pdf}} \textbf{Possible titles}: Write down ideas for possible titles for the visualization(s). 
\end{tightItemize}

These four rudder variants were the most common amongst participants in Study 1 who used writing, each mentioned by at least two participants. We focused on these specific variants to keep stimuli grounded in actual practice.
While this set is somewhat limited, it is well-suited to explore the impact of incorporating a step like this in the design process. 
These variants were directly based on discussions with the participants in Study 1, allowing us to remain within the scope of real-world experiences and avoid extrapolating to novel cases. As the nature of this study is exploratory, this approach helped maintain relevance and applicability to participants' existing workflows.

The interviewer introduced the variants in a random order, then walked through an example of what each rudder variant might look like in a sample design prompt. The order of examples was also randomized.
The interviewer asked if the participant had any questions on each variant or how it might be applied in the design process. They then asked the participant to reflect on all the variants discussed and to select the one they liked the best or found the most interesting. 

While a controlled study would have randomly assigned variants to participants, we prioritized external validity by allowing participants to choose their writing rudder, reflecting the flexibility of real-world design practices and ensuring higher engagement. This approach was fitting for the exploratory nature of this study, as it helped identify which rudders best aligned with existing design practices. However, this decision involved trade-offs. The lack of randomization limited our ability to directly compare each variant’s effectiveness across a representative sample. Additionally, participants' tendency to rate their chosen rudders favorably could be viewed as a result of wanting to give high ratings to the rudder they selected. However, participants chose rudders that they felt would work best for them, which also suggests that they had a prior preference for the rudder.

Following the introduction of these interventions, participants used their selected variant in a brief design exercise. This exercise was described via a pre-recorded video so all participants received the same information. Since the study focused on the beginning of the design process, the exercise was limited to 10 minutes. Participants were provided with one year of Chicago weather data: temperature, precipitation, and wind speed. They were asked to design a visualization or a set of visualizations for a marketing agency whose goal was to determine the optimal time to begin marketing a waterproof windbreaker. 

The template for the design exercise can be seen in \cref{fig:exercise_template}. Participants were provided the overall goal of the visualization, any specific client considerations they might need, and the audience for the visualization. This overview was set up to imitate the content received in initial conversations with clients (i.e., requirements gathering stage). If participants had other questions about the client's goals or interests (e.g., is there a particular temperature at which customers begin wearing windbreakers?), they were told to use their best judgment and to assume it aligned with what the client would say if asked. 

The ``writing step'' in the template was replaced with the participant's selected rudder variant. Depending on their selection, participants were prompted to generate three user questions (displayed in \cref{fig:exercise_template}), three possible reader conclusions, a 3-4 sentence story, or three possible titles for the visualization(s) they planned to make. 

Pilot testing indicated that designers felt under pressure when having their design process observed. Therefore, we opted not to observe the design process, since the exercise itself was also timed, and we wanted to limit the amount of external pressure placed on the participant. 
Participants were repeatedly assured that they were not expected to produce a final design, since this exercise was focused on the starting phases of visualization design. 
They could use any tools during the exercise and were not asked to share screen or narrate the process.

After 10 minutes had passed, the exercise ended and the camera was turned back on.
The interviewer asked about how the ``writing step'' affected the process of getting started on the design. The participant commented on the impact of the selected variant on their design process and considered the other three variants and what their impact may have been. The order of the three variants were introduced was randomized. 

Following the completion of the interview, participants took a 5-10 minute post-interview survey which asked them to further consider each of the four rudder variants and to rate their impact on the understand, ideate, and make stages of design from the DAF. They also rated the impact of the writing step on ``getting started on the design.'' This additional rating was added to the set of pre-defined stages to provide a more holistic evaluation of the rudders. By including ``getting started,'' we captured insights into the impact of writing rudders on how designers initiate the workflow, which could include a combination or non-linear progression of the stages from the DAF. 

Ratings were made on a scale from 1 to 5. A rating of 1 indicated that the variant would \textit{negatively} impact the stage in question, a rating of 3 indicated that the variant would have \textit{no impact} on their design process, and a rating of 5 indicated that the step would \textit{positively} impact the design process. All ratings were made in comparison to the participant's current design process. 
While this approach relied on subjective feedback, it provided valuable insights into designer perceptions of the proposed writing rudder variant's usefulness within existing workflows. This can help us identify potential challenges and areas where these new design approaches might integrate seamlessly.  
Participants also reported if they currently use a similar step (which did not have to be written) and if they would consider using a step like this in the future. At the end of the survey, they reported their overall industry of work and their typical design outcomes. 

In our discussion of themes from these interviews, participants are referred to with their assigned ID number [P\#]. To distinguish between studies, ID numbers for Study 2 begin at 101 rather than 1. 

\subsubsection{Participants}
\label{sec:study2_participants}

15 visualization designers participated in this study. Participants were recruited using the same calls for participation as in Study 1 (\href{https://www.datavisualizationsociety.org/}{DVS}, \href{https://newsnerdery.org/}{News Nerdery}, and social media posts) with the same eligibility requirements.
12 participants were employed full time, with two participants employed as students and one participant on leave. A majority of participants were women (13), with two men.
Further information on participants can be found in \cref{tab:study_participants} and supplementary materials.

\subsubsection{Coding}
\label{sec:study2_coding}

Interviews were again coded by the first two authors for the use of writing and the stage of the design process in which writing occurred (if any). The design outcome was captured in the post-interview survey.
For the codes where consensus was not found, the third author was brought in as a tiebreaker.
After considering responses from all three coders, consensus was reached on all coding categories.
Interrater reliability (Cohen's kappa) for the first two coders was 0.796 for the use of writing and 0.754 for the stage of the design process. 

In addition to these pre-defined codes, we also completed an open coding of the discussions about the rudder variants. The first two authors independently coded participant feedback on the variants, focusing on likes, dislikes, and the impact of these steps. This process of open coding allowed us to uncover themes and insights into how these steps influenced the design approach. After individual analysis, both authors met to discuss the emergent themes across the sets of codes.

\begin{table*}[!ht]
\centering
\def\arraystretch{1.1}
\begin{tabular}{lllR{0.68\linewidth}}
\hline
& & &\\[-1.75ex]
\textbf{ID} & \textbf{Industry} & \textbf{Process Start} & \textbf{Representative Quotes from Participants about the Selected Rudder}
\\ 
& & &\\[-1.75ex]
\hline & & &\\[-1.75ex]
\centeredmulticolumn{4}{figs/questions_icon.pdf}{Questions} \\
& & &\\[-1.75ex]
101& Research & Raw data   & ``You need an objective and a plan, and you need to make sure those questions are open enough.''
\\ 
103 & Public Sector & Sketch &  ``The establishing [of] the questions beforehand makes you sit down and just focus on the client first.''
\\ 
104 & Public Sector & Sketch & 
``Going back, saying, this is my goal... Can people answer this question?... I think that's super helpful.''
\\ 
106 & Healthcare & Raw data & 
``It's a non- event. It's just part of [design]... It's just pretty fundamental.'' \\ 
107 & Software & Tool & ``Building something that the user wants is the main goal. I think that doing [questions] is the most effective way.''
\\ 
113 & Research & Sketch & ``I think it's a good way to kind of organize things, cause I feel like a lot of times, it's just kind of in my head.''
\\ 
114 & Public Sector & Sketch &   ``I like this approach
and that it does require me to begin more with those [questions].''
\\ 
115 & Journalism & Raw data & ``I think it was a good framing to have in mind. But it definitely changed a lot as I like explored the data more.'' 
\\ 
\hline 
& & & \\ [-1.75ex]
\centeredmulticolumn{4}{figs/conclusion_icon.pdf}{Conclusions} \\
 & & &\\[-1.75ex]
102 & Journalism  & Sketch &  ``Coming up with what you want people to get out of this data set... helps me figure out what I'm gonna be visualizing'' \\ 
108 & Journalism & Tool &  ``Having to write out kind of the actual conclusion that someone would see forced me to really be strategic.''
\\ 
110 & Research & Tool &  ``I was surprised at how much it guided me in the process. Hadn't really occurred to me to to do it like that before.''
\\ 
111 & Technical & Writing   &   ``You want to make sure that you haven't gone down a rabbit hole too far, and you're straying from the main point.''
\\ 
112 & Research & Tool &  ``I would just have it as [a] starting point, because the takeaway can change.''
\\ 
\hline & & & \\[-1.75ex]
\centeredmulticolumn{4}{figs/story_icon.pdf}{Narrative} \\
 & & &\\[-1.75ex]
105 & Research & Sketch & ``It’s a good idea to try to put it into words… It helped to figure out what the point is.''
\\ 
109 & Public Sector  & Tool &  ``It really focused me. I used what I wrote to immediately start thinking about what graph type I was going to use.'' \\ 
\hline
\end{tabular}
\caption{Participant responses and quotes regarding the use of written rudders, grouped according to the  rudder type chosen by the participant. ``Process Start'' refers to how the participant usually begins their design process. Average ratings for  selected rudders of all participants were 4.25 or greater. All participants reported that they would use their selected variant in future contexts. 
}
 \label{tab:study2_quotes}
\end{table*}

\subsection{Results}
\label{sec:study2_results}

\subsubsection{Current Use of Writing}

As in Study 1, we coded each participant's use of written text and when in their typical design processes. Participants used text at these frequencies: deliberately (3/15), incidentally (5/15), and not at all (7/15).
A smaller proportion of participants (20\%) used written text deliberately in Study 2, compared to Study 1 (38\%). This may be partially due to the Study 2 interview, which asked fewer questions about the current use of written language than the Study 1 interview. 

As in Study 1, most participants (6/8) who used writing did so when \textit{understanding} the audience or goals of the visualization; a few (3/8) used language during the \textit{ideate} phase, when generating ideas for how to show the information. 
These findings support those from Study 1 as well as our decision to focus on the early stages of design in this study. 

\subsubsection{Intervention Results Overview}

Overall, participants responded positively to the written rudders, assigning an average rating of 3.71 out of 5 across the different variants and stages of design. Key questions received the highest average ratings, with almost all participants viewing this approach as beneficial to the design process. Across the rudder variants, all stages of the design process were rated relatively similarly on average, with the \textit{understand} stage rated highest (Mean = 3.80, SD = 1.1), followed by \textit{make} (Mean = 3.70, SD = 1.1), \textit{ideate} (Mean = 3.63, SD = 1.1), and getting started (Mean = 3.60, SD = 1.2). The average rating for each variant and stage of design can be seen in \cref{tab:avg_ratings}.

Two variants (questions and conclusions) received relatively positive average ratings across stages. Almost all participants (14/15) viewed the use of key questions as positive overall for the design process, with some emphasis on the \textit{understand} stage, as 9 participants reported a rating of a 5. There was also some emphasis on the \textit{make} stage, with 7 participants rating it a 5. Only one participant (P112) provided any neutral or negative ratings for the questions variant.

Most participants (14/15) also viewed writing possible conclusions as beneficial for the overall design process. 
Despite the overall positive reception, two participants rated the conclusion variant as having a negative effect on the \textit{ideate} stage and the process of getting started on the design. Four participants rated it as having no benefit for the design process, primarily for the \textit{make} (4/4) and \textit{ideate} (2/4) stages.

The other two variants (narrative and titles) did not fit well in the beginning of the design process; some participants felt that making a plan for the design too early could bias the exploration of the data. The narrative rudder variant elicited mixed opinions from participants, with it being just slightly more common that the rudder would have a positive effect (8/15) than a negative effect (6/15). While overall the impact was seen as more positive than negative, a majority of participants (10/15) gave the narrative rudder a neutral or negative rating for at least one of the stages of design, particularly the \textit{understand} stage (9/15) and getting started in the design overall (8/15). In other areas of the design process, the narrative rudder was seen relatively favorably, receiving high ratings for the \textit{make} (9/15) and \textit{ideate} (8/15) stages. This rudder was divisive for participants, but it may have a use case in the later stages of the design process when the visuals are being actively created.

The title variant received the most negative feedback of the four variants tested. More participants (6/15) felt it would hinder the design process than benefit it (4/15). Another third of participants felt it would have no benefit or adverse impact (5/15). Four unique participants gave a rating of a 1 to at least one stage of the design process. No other rudder variants received any `1' ratings for any other stage. Ratings of `1' and `2' for this rudder variant tended to be assigned most frequently to the act of getting started in the design process (9/15). 

The most popular rudder variant was the writing of key questions (8/15), followed by possible reader conclusions (5/15) and narratives (2/15). These selections and representative quotes from participants can be found in \cref{tab:study2_quotes}. These quotes provide additional detail on participant opinions on their selected rudders, which were generally positive. The commentary from participants reflects a variety of nuances as well, including considerations for using the rudder and more detail on the impact on the design process during the exercise. 

Most participants (11/15)  noted that there were similarities between the rudder variants, as ``they're all very similar, just approaching things at a different stage'' [P103]. This overlap stems from their common function, which is to guide the visualization by externalizing the message, story, or key goals of the design. While there were similarities between the variants, participants also displayed distinct thoughts about each one, as we explore in the themes.

Although the participants only used one of the variants during the design exercise, we report their thoughts and reflections on all four written rudders.  Some of these results are thus based on \textit{considering} the use of language in the hypothetical, while others are reflections from the \textit{actual} use of the rudder in the design exercise. 
When applicable, we provide this context.
Although we do not examine design outcomes due to the intentional focus on the design process and the limited timeframe, both hypothetical and actual reflections provide useful insights.

Participant responses to the writing step had an average length of 257 characters, ranging from 63 characters for the shortest response to 476 for the longest. Most responses (11/15) were concerned with time periods of extrema, particularly for precipitation (7/15), temperature (4/15) or wind (3/15), as relevant to the overall goal of the exercise in marketing waterproof windbreakers. Examples are shown in \cref{fig:teaser}.

The next subsections describe the results in more detail in the form of three themes. Data related to the  themes can be found in \cref{fig:study2_variants}. The precise impact of writing rudders may vary across different design contexts and tools, and these themes do not incorporate consideration of design outcomes. Additionally, we focused only on the initial stages of the design process. However, these themes were common across participants working in different industries with different backgrounds and tools and thus provide a useful starting point for understanding the effect of writing rudders within the scope of our study.

\subsubsection{Theme 1: Written Rudders Add Design Focus}
\label{sec:study2_theme1}

\begin{figure}
    \centering
    \includegraphics[width=0.9\linewidth]{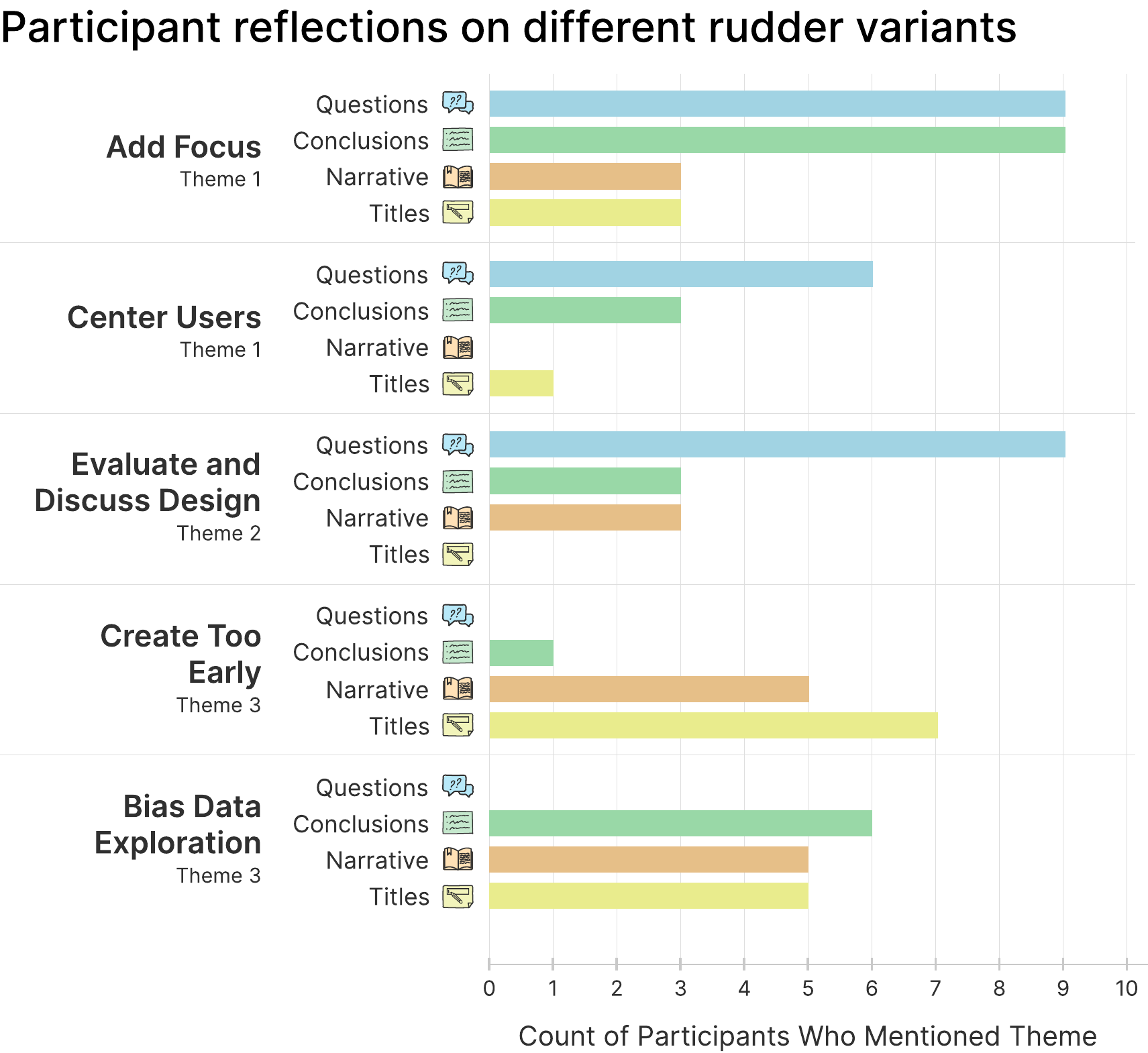}
    \caption{Participant reflections on the impact of different rudder variants in Study 2. Counts shown here represent how many participants (n = 15) mentioned the given feature for the specific variant. Questions and conclusions could add user-centered focus to the design. Some rudders may also be useful for later in the design process. One major concern was the potential to introduce bias.}
    \label{fig:study2_variants}
\end{figure}

When discussing the rudders, participants commented frequently on the structure and focus they added to the design process. These comments were mainly associated with the writing of key questions  (9/15) and the writing of example reader conclusions/takeaways (9/15). The narratives (3/15) and titles (3/15) were not often seen as providing additional direction to the design process, as in \cref{fig:study2_variants}.

In comparison to how participants typically began the design process, writing rudders allowed for a more guided process. The interviews revealed that in their current practices, participants began the process by entering the data into a visualization tool (5/15), examining the raw data (3/15), or sketching (6/15). Only one participant started their design process with writing. 
Through writing rudders, participants felt they had a greater degree of focus in the initial stages of the design process (12/15). For example, P108 said that writing out possible conclusions helped to narrow down, ``Which of the metrics would be most important to someone?''
Written rudders also acted as  guardrails to the design process, protecting against, ``getting too excited and diving into the data, potentially losing focus of what the purpose is.'' [P103].

In addition to acting as a guide when beginning the design process, written rudders emphasized a closer connection to the audience's perspective and their uses of the data (8/15). 
When describing why they most preferred the question approach, P107 mentioned, ``I like to make sure that I am doing what the audience wants... The data isn't valuable unless you're giving it to the right people in the right format.''
The narrative and title rudder variants did not seem to provide this user-centered guidance in the same way that questions or conclusions did.

While only hypothetical, future use of these written rudders was also received well. For both of the highest rated rudders (key questions and possible conclusions), almost all participants (14/15) reported that they would use the variant in the future in at least some contexts. 

In general, participants viewed the written rudders favorably, as they helped to provide additional focus and structure to the early stages of the design process. The listing of possible questions provided the most added benefit in bringing the design closer to the needs of the audience or user. 
Additional quotes regarding the design focus added by written rudders can be found in \cref{tab:study2_quotes}.

\subsubsection{Theme 2: Using Rudders for Evaluation or Instruction}
\label{sec:study2_theme2}

Participants also brought up the possible use of written rudders later in the design process (12/15).
These comments should be seen as hypothetical, since they pertain to stages of the design process that participants could not reach within the 10 minute exercise time frame. However, some reflections are directly connected to past client scenarios.

In providing a guide for beginning a visualization, written rudders create artifacts to act as a comparison for the final design. Participants (11/15) mentioned that the tangible outcome of writing rudders would be a useful metric with which to judge a design's success or to aid in discussions with a client. As shown in
\cref{fig:study2_variants}, the questions rudder was most frequently associated with possible evaluative use.

Along with the client or a potential user group, the designer could evaluate whether the resulting design measured up to the goals set out in the key questions (9/15) or  conclusions (3/15) outlined prior to beginning the design. Completing this step could help, ``at the end of the design process, to be like, wait a second. Are people actually coming away with what I wanted them to come away with?'' [P102]. 
A narrative (3/15) could also provide design justification for the client (``Definitely when you're communicating with a team or with the client... it would be nice to have a narrative'' [P105]) or simply to provide a point of engagement with the client about the visualization's goals.

In addition to supporting the client's goals for the visualization, written rudders could also be used in academic or educational contexts.
When describing their experience in writing out key questions, P110 described it as ``the more intuitive place to start for, particularly a beginner. But it's not a necessary step for me.'' 
The rudders themselves sometimes felt educational as well.  After completing the design exercise, P113 stated, ``I'm always a little bit skeptical of this sort of thing because I'm like, This feels like school. But I actually really liked it.''
Considering these reflections, there may be a tailored approach to using rudders that is best suited for students or early-career designers. 

The comments made by participants for this theme were not based in their direct experience using the rudders for evaluation or instruction; the design exercise was completed solo and only focused on the initial stages of the design process. However, reflecting on the rudder exercise and drawing from  prior design experiences, participants indicated that these features could have use cases beyond getting started on the design. 

\subsubsection{Theme 3: Rudders Suited for Later Stages}
\label{sec:study2_theme3}

While the narrative and title rudder variants were the least preferred by participants (as seen in \cref{fig:teaser}), participants  indicated that these variants could be better suited for later stages of the design process (9/15). When considering the narrative approach, P115 compared it to the ```alt text for graphic,'' stating, ``I can't imagine writing it before.'' 
In one example, P110 already used a narrative step as a rudder in their design process when selecting the visualization that best communicates the key ideas: ``I'll write text for each [visualization option]... I'll look at the text by itself to see which one reads better.'' 

As shown in \cref{fig:study2_variants}, writing titles before exploring the data was often seen as premature (7/15). Titles tended to be one of the last steps in the design process, and the title of a visualization may change over the course of the design. Completing this step at the beginning of the design process seemed counterintuitive. 
Shifting some rudder variants (conclusions, narratives, and possible titles) to later in the design process has additional benefits for data exploration. While written rudders provided a clear direction to the design process, this design direction guide could also introduce possible bias to data exploration (10/15).
When describing the impact of the rudder on their design process, P112 mentioned, ``[the design] was 100\% based on on on these takeaways.'' 

This kind of direction could be useful when focusing on explaining data, but some visualization contexts require unbiased data exploration. F
For P102 (journalist), ``if there's no lead, there's no story... It's a very easy test that we really have to be able to answer.'' In this context, writing out possible takeaways or titles could ensure data relevance. On the other hand, P106 (researcher) described, ``For me, quality data vis means you don't know the answers. You're exploring the data.'''

Participants suggested workarounds for the potential issue of bias, such as using a ``fill in the blank'' approach rather than referring to specifics of the data. An example of this approach is shown for the conclusions rudder in  \cref{fig:teaser}, using an X instead of a specific month. 
Keeping an open mind would be important when using these approaches, as mentioned by P108, `` I do think it is helpful to be imagining what someone looking at [the visualization] might be like experiencing or thinking, but, on the other hand, I think you also have to be open to the idea of the data not saying what you might want it to say.''

\section{Discussion}
\label{sec:overall_discussion}

In exploring the role of writing in visualization design, these two interview studies offer a nuanced understanding of how written rudders can potentially enrich current visualization design practices.
Overall, participants in Study 2 responded positively to the use of rudders when designing visualizations. All participants indicated that they would be willing to adopt some form of this idea, despite the fact that only a few (3/15) use writing in their current practice. 
In particular, writing down key questions and possible  conclusions emerged as preferred variants. 

\textbf{The implementation of writing rudders into the design workflow has the potential to improve the clarity and focus of visualization design. }
When reviewing the rudder variants in Study 2, designers often made comments that they used similar steps, ``I do use a combination of these throughout the process'' [P107]. 
However, these steps happened mostly internally (mentally).
By explicitly combining language and visualization, rudders help bridge the gap between the data and its interpretation, facilitating a deeper connection with the audience and a better understanding of the design process goals.
Creating detailed plans or task lists can improve outcomes, specifically in the cases of writing tasks or employee performance \cite{kaur2018creating, gollwitzer1996psychology, parke2018daily}. Research suggests that this works  in part by increasing engagement with the overall objective. 

While traditional design practices, such as writing down design requirements or documenting decision rationale \cite{munzner2009nested, walny2019data, mckenna2014design}, are well-established in the field, writing rudders provided an additional layer of user-centered guidance  \cite{norman2013design, vredenburg2002survey} to the design process. 
Compared to other design practices, such as cognitive walkthroughs, the addition of written rudders takes less time (less than 10 minutes) and effort.
Written rudders also provided a narrative focus, crucial for creating engaging and dynamic designs to effectively tell stories with data. Rudders build on work in narrative visualization, offering a lightweight and flexible approach to encourage narrative designs \cite{segel2010narrative, hullman2011visualization}.

Furthermore, writing rudders can also act as artifacts that aid in evaluating a design’s success and facilitating continuous iteration. Similar to model checking and graphical belief elicitation methods \cite{kale2024evm, mahajan2022vibe, kim2017explaining}, rudders prompt designers to externalize assumptions and consider the audience’s perspective. This approach may not only aid in the initial stages of understanding and ideating but also may support ongoing evaluation and refinement. The possible use of rudders throughout the design process further supports them an interesting case study in how to incorporate written language into the visualization design process.

\textbf{While writing rudders may have important uses and benefits to visualization design practices, there are also several considerations for future use.} 
In some cases, designers want to avoid any preconceived notions that may bias data exploration. 
To avoid bias in data exploration, designers could use \textit{only} a series of open-ended questions to explore key pieces of the data without stating expectations of the data features. Designers could alternatively use an process similar to hypothesis testing. A combination of writing and sketching could provide an a priori understanding of how data features would look if different possible takeaways were true \cite{mahajan2022vibe, kim2017explaining}. 

The benefits and drawbacks of writing rudders may also vary based on the designer's role or stage of the design process. Using a rudder maybe more beneficial for early career designers or in educational settings, rather than in established workflows. In collaborative settings, rudders may help align team objectives but could also limit creative exploration. Certain variants, such as the narrative, may be more useful once a draft of the design has been created. These considerations point to important directions of future work on written rudders.

\section{Limitations and Future Work}
\label{sec:limitations_futurework}

\pheading{Need for broader controlled studies}:
One significant limitation of this study is the absence of controlled studies to statistically evaluate the effectiveness of writing rudders against other methods or tools. While our preliminary findings suggest that the rudder can provide additional focus and clarity to the design process, these results are not indicative that this is the best approach available. Additionally, this study was focused on a specific set of written rudders used in the early stages of the design process, thus limiting the generalizability of the findings.

Our current study serves as an initial exploration of designer perspectives on the rudder intervention, providing groundwork on considerations that should be made for future variants and evaluations.
Future research should include controlled user studies that compare rudders with established design methods, using metrics such as efficiency, creativity, and outcome quality. This work could also incorporate novel variants of written rudders and examine different stages of the design process. This would provide the necessary statistical evidence and broader scope to substantiate the rudder’s practical value. 

\pheading{Reflections and hypotheticals}: 
The primary goal of this paper was to uncover the current  use of writing in visualization design, and ascertain practitioners' reaction to written rudders as a part of the design process. 

Participant feedback is based on reflections and hypothetical scenarios rather than long-term application of the steps in regular workflows. 
This approach, while useful for forming an initial understanding, may not capture the complexity of actually using these practices consistently. Preferences expressed in hypothetical scenarios may not translate into actual changes in established design practices. 

Future research should transition from hypothetical scenarios to empirical evaluations, including the use of longitudinal studies to observe how designers integrate written rudders into their workflows over time. By observing the adoption of these practices in a variety of real-world design projects, researchers can gather concrete evidence on the utility of written rudders, including barriers to integration. 

\pheading{Leaving out the design}:
This paper focused specifically on the design \textit{process}, and the interviews were primarily about the early stages of creating a visualization. This focus excludes the direct impact on the design outcomes. This study was not designed to assess outcomes and instead focused on the impact on the design \textit{process}. This is a preliminary study on interventions like writing rudders to check for viability and overall designer interest. In pilot interviews for Study 2, designers mentioned that observing their design process or evaluating their design outcome put a great deal of pressure on the design exercise. To encourage participation and to focus the interview on the rudders themselves, rather than on unrelated design decisions, we opted for a study design that did not include evaluation of design outcomes.

Subsequent research should directly investigate how writing rudders impact the final designs. While it is useful to understand their impact on the design process, it is unclear whether these impacts have any downstream effects on the resulting design. By comparing designs developed with and without these interventions, researchers can better quantify their impact on creativity, clarity, user engagement, and the use of text in the design. Evaluations of design outcomes will allow for a more concrete assessment of the benefits and limitations of written interventions and is an essential next step in building a better understanding of the impact of writing in the visualization design process.

\section{Conclusion}
\label{sec:conclusions}
This research has explored the role of writing in the visualization design process, revealing its potential to enhance clarity and intention during the creation of a visualization as well as to provide artifacts for evaluating designs at the end of the design process. Our findings suggest that by incorporating written rudders, such as formulating key questions or possible reader takeaways, designers can not only refine their goals for the design but also align more closely with audience needs. Future studies should explore the longitudinal impact of these practices on design processes and outcomes. Overall, these results reinforce the symbiotic relationship between language and visualizations, encouraging further integration of the two.


\section*{Supplemental Materials}
\label{sec:supplemental_materials}

All supplemental materials are available at \href{https://osf.io/yjsnh/?view\_only=12cda64e58994f5b81b001328368dd49}{OSF}, released under a CC BY 4.0 license.
In particular, they include (1) interview materials for both studies and (2) analysis files used for participant distributions and exploratory data analysis.


\acknowledgments{%
  The authors wish to thank Bridget Cogley for initial feedback on the interview materials and Simone Laszuk for feedback on figures and manuscript clarity.
  This work was supported by the National Science Foundation Graduate Research Fellowship under Grant No. DGE 2146752.
}

%

\bibliographystyle{abbrv-doi-hyperref-narrow}

\bibliography{references}

\appendix 

\end{document}